\documentclass[review]{elsarticle}

\usepackage{lineno}
\usepackage{amsfonts}
\usepackage{amssymb}
\usepackage{amstext}
\usepackage{graphicx}
\usepackage{amsmath}
\usepackage{url}
\usepackage{epstopdf}
\usepackage{color}
\usepackage{xcolor}
\usepackage{bbm}
\usepackage{wrapfig}
\usepackage{lscape}
\usepackage{rotating}

\usepackage{array}
\usepackage{subfig}

\modulolinenumbers[5]

\journal{Journal of Information Fusion}

\graphicspath{{./figures/}}

\begin{document}

\begin{frontmatter}

\title{\textbf{A Message Passing Approach for Decision Fusion in Adversarial Multi-Sensor Networks}}


\author[mymainaddress]{Andrea Abrardo, Mauro Barni, Kassem Kallas and Benedetta Tondi}


\address[mymainaddress]{Department of Information Engineering and Mathematics, Via Roma 56, Siena, Italy}

\begin{abstract}

We consider a simple, yet widely studied, set-up in which a Fusion Center (FC) is asked to make a binary decision about a sequence of system states by relying on the possibly corrupted decisions provided by byzantine nodes, i.e. nodes which deliberately alter the result of the local decision to induce an error at the fusion center. When independent states are considered, the optimum fusion rule over a batch of observations has already been derived, however its complexity prevents its use in conjunction with large observation windows.

In this paper, we propose a near-optimal algorithm based on message passing that greatly reduces the computational burden of the optimum fusion rule. In addition, the proposed algorithm retains very good performance also in the case of dependent system states. By first focusing on the case of small observation windows, we use numerical simulations to show that the proposed scheme introduces a negligible increase of the decision error probability compared to the optimum fusion rule. We then analyse the performance of the new scheme when the FC make its decision by relying on long observation windows. We do so by considering both the case of independent and Markovian system states and show that the obtained performance are superior to those obtained with prior suboptimal schemes. As an additional result, we confirm the previous finding that, in some cases, it is preferable for the byzantine nodes to minimise the mutual information between the sequence system states and the reports submitted to the FC, rather than always flipping the local decision.

\end{abstract}


\end{frontmatter}


\section{Introduction}

Decision fusion for distributed  detection has received an increasing attention for its importance in several applications, including  wireless
networks, cognitive radio, multimedia forensics and many others. One of the most common scenarios is the parallel distributed fusion model.  According to this model, the $n$ nodes of a multi-sensor network gather information about a system and make a local decision about  the system status. Then the nodes send the local decisions to a Fusion Center (FC), which is in charge of making a final decision about the state of the system. \cite{Vemp13}

In this paper, we focus on an adversarial version of the above problem,  in which a number of malicious nodes, often referred to as Byzantines \cite{Vemp13}, aims at inducing a decision error at the FC \cite{WSNDDByz}. This is a recurrent problem in many situations wherein the nodes may make a profit from a decision error. As an example, consider a cognitive radio system \cite{WLSH10,Raw11,Zhang2013secure,wang2014secure} in which secondary users cooperate in sensing the frequency spectrum to decide about its occupancy and the possibility to use the available spectrum to transmit their own data. While cooperation among secondary users allows to make a better decision, it is possible that one or more users deliberately alter their measurements to let the system think that the spectrum is busy, when in fact it is not, in order to gain an exclusive opportunity to use the spectrum. Online reputation systems offer another example \cite{Sun12}. Here a fusion center must make a final decision about the reputation of an item like a good or a service by relying on user's feedback. Even in this case, it is possible that malevolent users provide a fake feedback to alter the reputation of the item under inspection. Similar examples are found in many other applications, including wireless sensor networks \cite{WSNDDByz}, \cite{WLSH10}, distributed detection \cite{Mar09}, \cite{DistrDetTree}, multimedia forensics \cite{Bar13} and adversarial signal processing \cite{AdvSP}.

In this paper we focus on a binary version of the fusion problem, wherein the system can assume only two states. Specifically, the nodes observe the system over $m$ time instants and make a local decision about the sequence of system states. Local decisions are not error-free and hence they may be wrong with a certain error probability. Honest nodes send their decision to the fusion center, while byzantine nodes try to induce a decision error and hence flip the local decision with probability $P_{mal}$ before sending it to the FC. The fusion center knows that some of the nodes are Byzantines with a certain probability distribution, but it does not know their position.

\subsection{Prior Work}

In a simplified version of the problem, the FC makes its decision on the status of the system at instant $j$ by relying only on the corresponding reports, and ignoring the node reports relative to different instants. In this case, and in the absence of Byzantines, the Bayesian optimal fusion rule has been derived in \cite{OptFusion},\cite{Var97} and it is known as Chair-Varshney rule. If local error probabilities are symmetric and equal
across the network, Chair-Varshney rule boils down to simple majority-based decision. In the presence of Byzantines, Chair-Varshney rule requires the knowledge of Byzantines' positions along with the flipping probability $P_{mal}$. Since this information is rarely available, the FC may resort to a suboptimal fusion strategy.

In \cite{Mar09}, by adopting a Neyman-Pearson setup and assuming that the byzantine nodes know the true state of the system, the asymptotic performance obtainable by the FC are analysed as a function of the percentage of Byzantines in the network. By formalising the attack problem as the minimisation of the Kullback-Leibler distance between the reports received by the FC under the two hypotheses, the blinding percentage, that is, the percentage of Byzantines irremediably compromising the possibility of making a correct decision, is determined.

In order to improve the estimation of the sequence of system states, the FC can gather a number of reports provided by the nodes before making a global decision (multiple observation fusion). In cooperative spectrum sensing, for instance, this corresponds to collectively decide about the white holes over a time window, or, more realistically, at different frequency slots. The advantage of deciding over a sequence of states rather than on each single state separately, is that in such a way it is possible for the FC to understand which are the byzantine nodes and discard the corresponding observations (such an operation is usually referred to as Byzantine isolation). Such a scenario has also been studied in \cite{Mar09}, showing that - at least asymptotically - the blinding percentage is always equal to 50\%. In \cite{VarshneyHardIso}, the analysis of \cite{Mar09} is extended to a situation in which the Byzantines do not know the true state of the system. Byzantine isolation is achieved by counting the mismatches between the reports received from each node and the global decision made by the FC. The performance of the proposed scheme are evaluated in a cognitive-radio scenario for finite values of $n$. In order to cope with the lack of knowledge about the strategy adopted by the attacker, the decision fusion problem is casted into a game-theoretic formulation, where each party makes the best choice without knowing the strategy adopted by the other party.

A slightly different approach is adopted in \cite{LearnByzantines}. By assuming that the FC is able to derive the statistics of the reports submitted by honest nodes, Byzantine isolation is carried out whenever the reports received from a node deviate from the expected statistics. In this way, a correct decision can be made also when the percentage of Byzantines exceeds 50\%. The limit of the approach proposed in \cite{LearnByzantines}, is that it does not work when the reports sent by the Byzantines have the same statistics of those transmitted by the honest nodes. This is the case, for instance, in a perfectly symmetric setup with equiprobable system states, symmetric local error probabilities, and an attack strategy consisting of simple decision flipping.

A soft isolation scheme is proposed in \cite{CDC14}, where the reports from suspect byzantine nodes are given a lower importance rather being immediately discarded. Even in \cite{CDC14}, the lack of knowledge at the FC about the strategy adopted by the attacker (and viceversa) is coped with by adopting a game-theoretic formulation.
A rather different approach is adopted in \cite{tolerant_scheme}, where a tolerant scheme that mitigates the impact of Byzantines on the global decision is used rather that removing the reports submitted by suspect nodes from the fusion procedure.

When the value of $P_{mal}$ and the probability that a node is Byzantine are known, the optimum fusion rule under multiple observation can be derived \cite{TIFSDFByz}. Since $P_{mal}$ is usually not known to the FC, in \cite{TIFSDFByz} the value of $P_{mal}$ used to define the optimum fusion rule and the value actually used by the Byzantines are strategically chosen in a game-theoretic context. Different priors about the distribution of Byzantines in the network are considered ranging from an extreme case in which the exact number of Byzantines in the network is known to a maximum entropy case. One of the main results in \cite{TIFSDFByz} is that the best option for the Byzantines is not to always flip the local decision (corresponding to $P_{mal} = 1$), since this would ease the isolation of malicious nodes. In fact, for certain combinations of the distribution of Byzantines within the network and the length of the observation window, it is better for the Byzantines to minimise the mutual information between the reports submitted to the FC and the system states.

\subsection{Contribution}

The main problem of the optimum decision fusion scheme proposed in \cite{TIFSDFByz} is its computational complexity, which grows exponentially with the length of the observation window. Such a complexity prevents the adoption of the optimum decision fusion rule in many practical situations. Also the results regarding the optimum strategies of the Byzantines and the FC derived in \cite{TIFSDFByz} refer only to the case of small observation windows.

In the attempt to diminish the computational complexity while minimising the loss of performance with respect to the optimum fusion rule, we propose a new, nearly-optimum, fusion scheme based on message passing and factor graphs. Message passing algorithms, based on the so called Generalised Distributive Law (GLD, \cite{genlaw},\cite{genlawnewlook}), have been widely applied to solve a large range of optimisation problems, including decoding of Low Density Parity Check (LDPC) codes \cite{GallagerLDPC} and BCJR codes \cite{genlaw}, dynamic programming \cite{verdu1987OptwithMP}, solution of probabilistic inference problems on Bayesian networks \cite{beliefpropagationAI} (in this case message passing algorithms are known as {\em belief propagation}). Here we use message passing to introduce a near-optimal solution of the decision fusion problem with multiple observation whose complexity grows only linearly with the size of the observation window, thus marking a dramatic improvement with respect to the exponential complexity of the optimal scheme proposed in \cite{TIFSDFByz}.

Using numerical simulations and by first focusing on the case of small observation windows, for which the optimum solution can still be applied, we prove that the new scheme gives near-optimal performance at a much lower complexity than the optimum scheme. We then use numerical simulations to evaluate the performance of the proposed method for long observation windows. As a result, we show that, even in this case, the proposed solution maintains the performance improvement over the simple majority rule, the hard isolation scheme in \cite{VarshneyHardIso} and the soft isolation scheme in \cite{CDC14}.

As opposed to previous works, we do not limit our analysis to the case of independent system states, but we extend it to a more realistic scenario where the sequence of states obey a Markovian distribution \cite{HMMref} as depicted in Figure \ref{fig.HMM}. The Markovian model is rather common in the case of cognitive radio networks \cite{CRMM1, CRMM2, CRMM3} where the primary user occupancy of the spectrum is often modelled  as a Hidden Markov Model (HMM).
%
%
The Markovian case is found to be more favourable for the FC with respect to the case of independent states, due the additional a-priori information available to the FC in this case.

Last but not the least, we confirm that the dual optimum behaviour of the Byzantines observed in \cite{TIFSDFByz} is also present in the case of large observation windows, even if in the Markovian case, the Byzantines may continue using the maximum attack power ($P_{mal} = 1$) for larger observation windows.

The rest of this paper is organised as follows. In Section \ref{sec:Notations}, we introduce the notation used in the paper and give a precise formulation of the addressed problem. In Section \ref{sec:MessagePassing}, we describe the new message passing decision rule based on factor graph. In Section \ref{sec:Simulations}, we first discuss the complexity of the proposed solution compared to the optimal solution. Then, by considering both independent and Markovian system states, we compare the performance of the message passing algorithm to the majority rule, the hard isolation scheme \cite{VarshneyHardIso}, the soft isolation scheme described in \cite{CDC14} and the optimal fusion rule. In addition, we discuss the impact that the length of the observation window has on the optimal behaviour of the Byzantines. We conclude the paper in Section \ref{sec:conclusion} with some final remarks.

\section{Notation and Problem Formulation} \label{sec:Notations}
\begin{figure}[t!]
\centering
    \includegraphics[width=1.0\textwidth , height = 6.5cm]{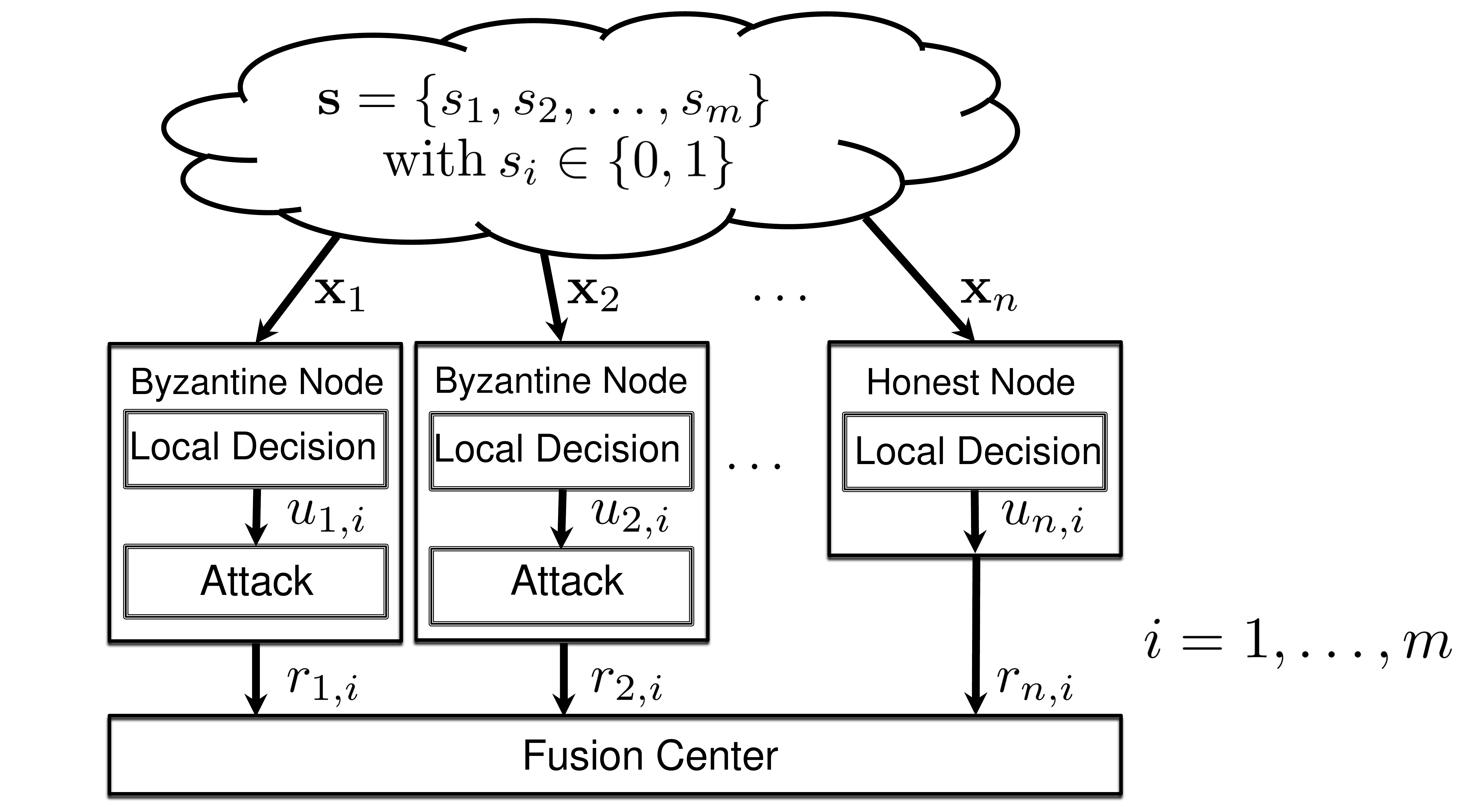}
    \caption{Sketch of the adversarial decision fusion scheme.}
    \label{fig.setup}
\end{figure}

%

The problem faced with in this paper, is depicted in Figure \ref{fig.setup}. We let $\mathbf{s} = \left\{s_1,s_2,\ldots,s_m\right\}$ with $s_i \in \{0,1\}$ indicate the sequence of system states over an observation window of length $m$. The nodes collect information about the system through the vectors ${\bf x}_1, {\bf x}_2 \dots {\bf x}_n$, with ${\bf x}_j$ indicating the observations available at node $j$. Based on such observations, a node $j$ makes a local decision $u_{i,j}$ about system state $s_i$. We assume that the local error probability, hereafter indicated as $\varepsilon$, does not depend on either $i$ or $j$. The state of the nodes in the network is given by the vector $\mathbf{h} = \left\{h_1,h_2,\ldots,h_n\right\}$ with $h_j = 1/0$ indicating that node $j$ is honest or Byzantine, respectively. Finally, the matrix $\mathbf{R} = \left\{r_{i,j}\right\}$, $i = 1,\ldots,m$, $j = 1,\ldots,n$ contains all the reports received by the FC. Specifically, $r_{i,j}$ is the report sent by node $j$ relative to $s_i$. As stated before, for honest nodes we have $u_{i,j}= r_{i,j}$ while, for Byzantines we have $p (u_{i,j} \ne r_{i,j} ) = P_{mal}$. The Byzantines corrupt the local decisions independently of each other.

\begin{figure}[t!]
\centering
    \includegraphics[width=0.6\textwidth]{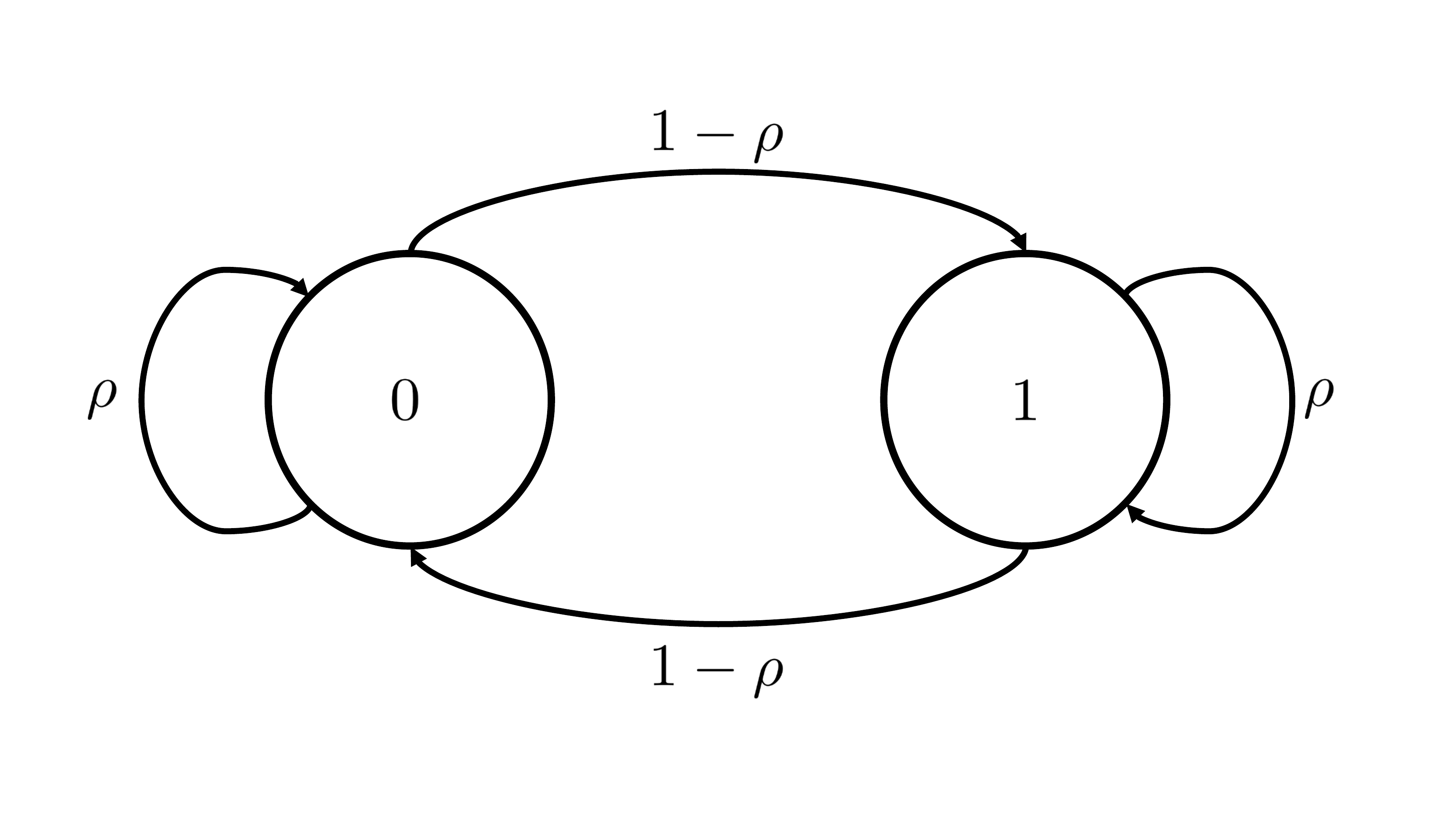}
    \caption{Markovian model for system states. When $\rho = 0.5$ subsequent states are independent.}
    \label{fig.HMM}
\end{figure}
By assuming that the transmission between nodes and fusion center takes place over error-free channels, the report is equal to the local decision with probability 1 for honest nodes and with probability $1-P_{mal}$ for Byzantines. Hence, according to the local decision error model, we can derive the probabilities of the reports for honest nodes:

\begin{equation}
p\left(r_{i,j} | s_i, h_j = 1\right) = (1-\varepsilon)\delta(r_{i,j}-s_i)+\varepsilon (1-\delta(r_{i,j}-s_i)),
\label{eqerr1}
\end{equation}

where $\delta(a)$ is defined as:
\begin{equation}
    \delta(a)=
    \begin{cases}
      1, & \text{if}\ a=0 \\
      0, & \text{otherwise}.
    \end{cases}
\label{delta}
\end{equation}

On the other hand, by introducing $\eta = \varepsilon(1-P_{mal}) + (1-\varepsilon)P_{mal}$, i.e., the probability that the fusion center receives a wrong report from a byzantine node, we have:

\begin{equation}
p\left(r_{i,j} | s_i, h_j = 0\right) = (1-\eta)\delta(r_{i,j}-s_i)+\eta (1-\delta(r_{i,j}-s_i))
\label{eqerr2}
\end{equation}

%

As for the number of Byzantines, we consider a situation in which the states of the nodes are independent of each other and the state of each node is described by a Bernoulli random variable with parameter $\alpha$, that is $p(h_j=0)=\alpha , \forall j$. In this way, the number of byzantine nodes in the network is a random variable following a binomial distribution, corresponding to the maximum entropy case \cite{TIFSDFByz} with  $p\left(\mathbf{h}\right) = \prod \limits_{j} p(h_j)$, where $p(h_j) = \alpha(1-h_j) + (1-\alpha)h_j$.

Regarding the sequence of states $\mathbf{s}$, we assume a Markov model as shown in Figure \ref{fig.HMM} , i.e., $p\left(\mathbf{s}\right) = \prod \limits_{i} p(s_i|s_{i-1})$. The transition probabilities are given by $p(s_i|s_{i-1}) = 1-\rho$ if $s_i = s_{i-1}$ and $p(s_i|s_{i-1}) = \rho$ when $s_i \ne s_{i-1}$, whereas for $i = 1$ we have $p(s_1|s_{0}) = p(s_1) = 0.5$.

In this paper we look for the the \emph{bitwise} Maximum A Posteriori Probability (MAP) estimation of the system states $\left\{s_i\right\}$ which reads as follows:


\begin{equation}
\begin{array}{cccc}
\hat{s}_i & = & \arg \max \limits_{s_i \in \{0,1\}} ~ p\left(s_i | \mathbf{R}\right) & \\
 & = & \mathop{\arg \max}\limits_{s_i \in \{0,1\}} \sum \limits_{ \{\mathbf{s},\mathbf{h}\} \backslash s_i} p\left( \mathbf{s}, \mathbf{h} | \mathbf{R}\right) & \textrm{(law of total probability)}\\
 & = &\mathop{\arg \max}\limits_{s_i \in \{0,1\}} \sum \limits_{  \{\mathbf{s},\mathbf{h}\} \backslash s_i} p\left(\mathbf{R} | \mathbf{s}, \mathbf{h} \right) p (\mathbf{s}) p (\mathbf{h}) & \textrm{(Bayes)}\\
 & = & \mathop{\arg \max}\limits_{s_i \in \{0,1\}} \sum \limits_{  \{\mathbf{s},\mathbf{h}\} \backslash s_i} \prod \limits_{i,j}p\left(r_{i,j} | s_i, h_j \right)\prod \limits_{i} p(s_i|s_{i-1}) \prod \limits_{j} p(h_j)  &
\end{array}
 \label{eqNN10}
\end{equation}

where the notation $\sum\limits_{\backslash}$ denotes a summation over all the possible combinations of values that the variables contained in the expression within the summation may assume by keeping the parameter listed after the operator $\backslash$ fixed. For a given $\mathbf{h}$, the matrix of the observations $\mathbf{R}$ at the FC follows a HMM \cite{ephraim2002hidden}. 
The optimisation problem in \eqref{eqNN10} has been solved in \cite{TIFSDFByz} for the case of independent system states. Even in such a simple case, however, the complexity of the optimum decision rule is exceedingly large, thus limiting the use of the optimum decision only in the case of small observation windows (typically $m$ not larger than 10). In the next section we introduce a sub-optimum solution of \eqref{eqNN10} based on message passing, which greatly reduces the computational complexity at the price of a negligible loss of accuracy.



\section{A Decision Fusion Algorithm Based on Message Passing} \label{sec:MessagePassing}
\subsection{Introduction to Sum-product message passing} \label{sec:MessagePassing Intro}
In this section we provide a brief introduction to the message passing (MP) algorithm for marginalization of sum-product problems. Let us start by considering $N$ binary
variables $\mathbf{z} = \{z_1,z_2,\ldots,z_N\}$, $z_i \in\{0,1\}$. Then, consider the
function $f\left(\mathbf{z}\right)$ with factorization:
\begin{equation}
f\left(\mathbf{z}\right) = \prod\limits_{k}f_k\left(\mathcal{Z}_k\right)\label{eq1}
\end{equation}
where $f_k$, $k = 1,\hdots,M$ are functions of a subset $\mathcal{Z}_k$ of the
whole set of variables.
We are interested in computing the marginal of $f$ with respect to a general variable $z_i$, defined as the sum of $f$ over all possible values of $\mathbf{z}$, i.e.:
\begin{equation}
\mu(z_i) = \sum\limits_{\mathbf{z}\backslash z_i}
\prod\limits_{k}f_k\left(\mathcal{Z}_k\right)\label{eq2}
\end{equation}
where notation $\sum\limits_{\mathbf{z}\backslash z_i}$ denotes a
sum over all possible combinations of values of the variables in $\mathbf{z}$ by keeping $z_i$ fixed. Note that marginalization problem occurs when we want to compute any arbitrary probability from joint probabilities by summing out variables that we are not interested in. In this general setting, determining the marginals by
exhaustive search requires $2^N$ operations. However, in many situations it is possible to exploit the distributive law of multiplication to get a substantial reduction in complexity.\\
To elaborate, let associate with problem
(\ref{eq2}) a bipartite \emph{factor graph}, in which for each variable we draw a variable node (circle) and for each function we draw a
factor node (square). A variable node is connected to a factor node $k$ by an edge if and only if the corresponding variable belongs to $\mathcal{Z}_k$. This means that the set of vertices is partitioned into two groups (the set of nodes
corresponding to variables and the set of nodes corresponding to factors) and that
an edge always connects a variable node to a factor node. \\
Let now assume that the factor graph is a
single tree, i.e., a connected graph where there is an unique path
to connect two nodes.
In this case, it is straightforward to derive an algorithm which
allows to solve the marginalization problem with reduced complexity. The algorithm is the
MP algorithm, which has been broadly used in the last years in
channel coding applications \cite{David}, \cite{Abr1}.

\begin{figure}[ptb]
\begin{center}
\includegraphics[width = 8cm]{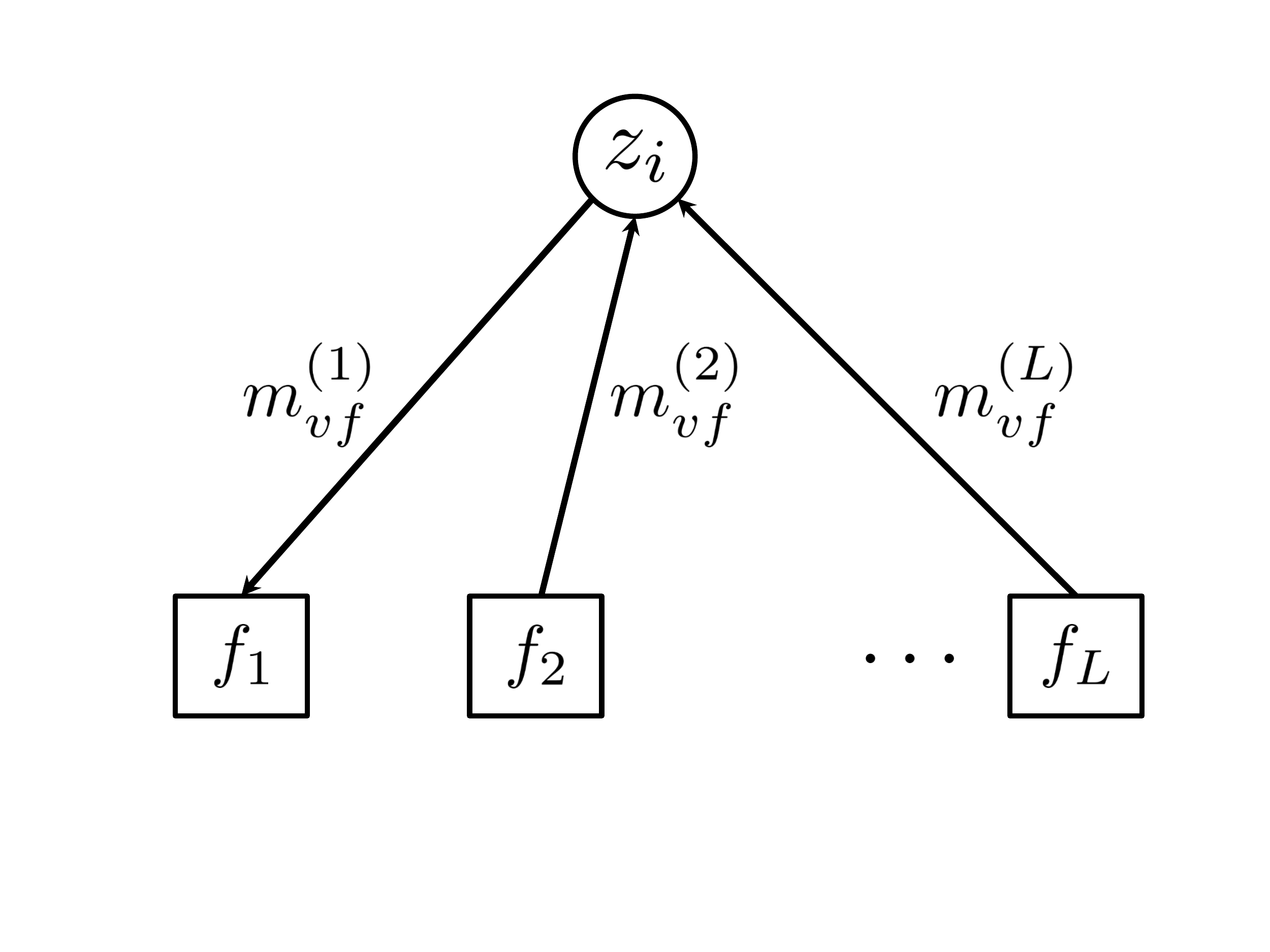}
\caption{Node-to-factor message passing.}%
\label{Fig1}%
\end{center}
\end{figure}

\begin{figure}[ptb]
\begin{center}
\includegraphics[width = 8cm]{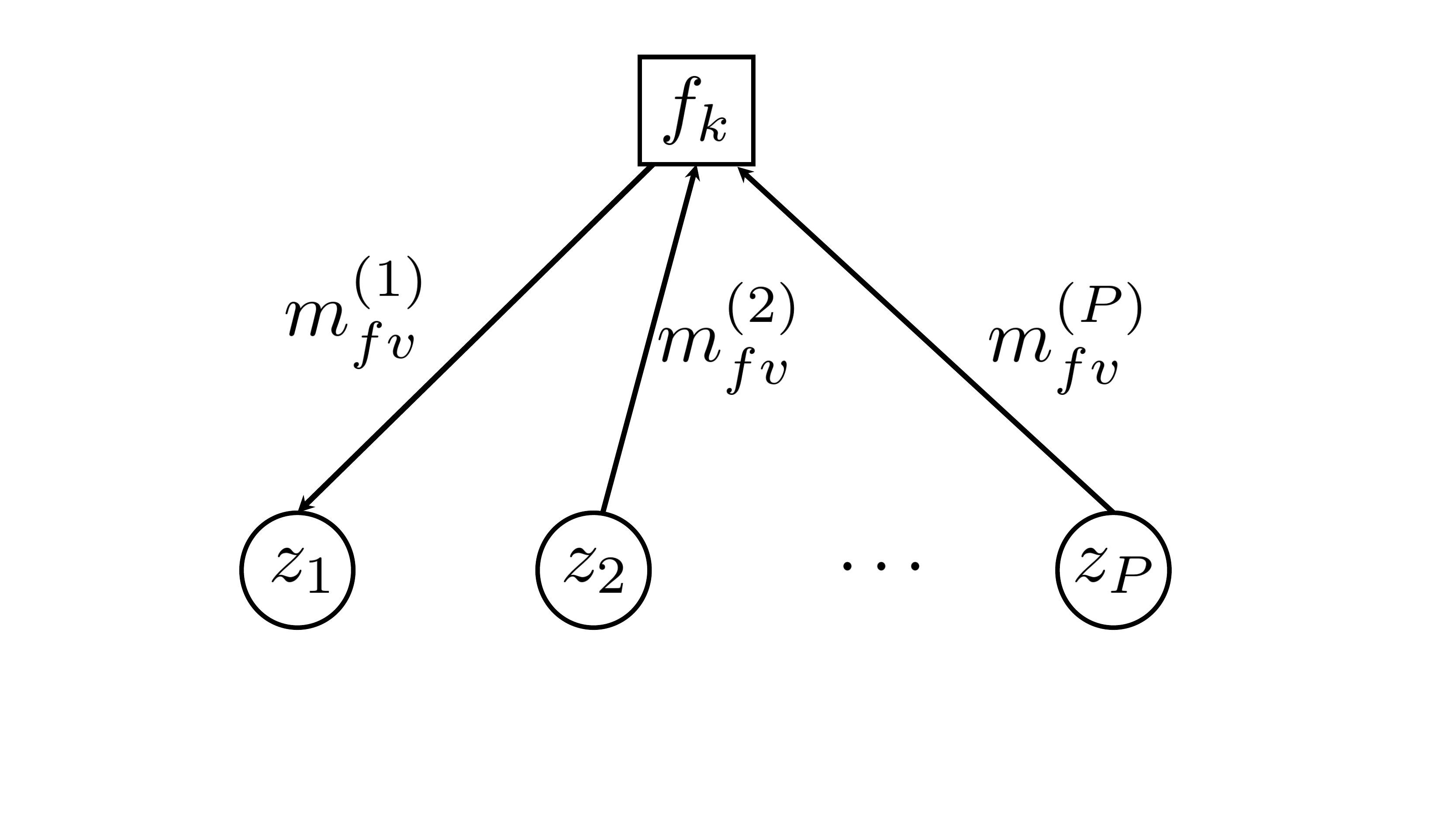}
\caption{Factor-to-node message passing.}%
\label{Fig2}%
\end{center}
\end{figure}

To describe how the MP algorithm works, let us first define messages as $2$-dimensional vectors, denoted
by $\mathbf{m} = \left\{m(0),m(1)\right\}$. Such messages are exchanged between variable nodes and function nodes and viceversa, according to the following rules.
Let us first consider variable-to-function messages ($\mathbf{m}_{vf}$), and take the portion of factor graph depicted in Fig. \ref{Fig1} as an illustrative example. In this graph, the variable node $z_i$ is connected to $L$ factor
nodes, namely $f_1,f_2,\ldots,f_L$. For the MP algorithm
to work properly, node $z_i$ must deliver the messages
$\mathbf{m}^{(l)}_{vf}$, $l = 1,\ldots,L$ to all its adjacent nodes.
Without loss of generality, let us focus on message
$\mathbf{m}^{(1)}_{vf}$. Such a message can be evaluated and
delivered upon receiving messages $\mathbf{m}^{(l)}_{fv}$, $l =
2,\ldots,L$, i.e., upon receiving messages from all function nodes
except $f_1$. In particular, $\mathbf{m}^{(1)}_{vf}$ may be
straightforwardly evaluated by calculating the element-wise product of
the incoming messages, i.e.:
\begin{equation}
{m}^{(1)}_{vf}(q) = \prod\limits_{j=2}^{L} {m}^{(j)}_{fv}(q)
\label{eq4}
\end{equation}
for $q=0,1$.
Let us now consider factor-to-variable messages, and refer to the
factor graph of Fig. \ref{Fig2} where $P$ variable nodes are
connected to the factor node $f_k$, i.e., according to the previous
notation, $\mathcal{Z}_k = \{z_1,\ldots,z_P\}$. In this case, the node $f_k$
must deliver the messages $\mathbf{m}^{(l)}_{fv}$, $l = 1,\ldots,P$ to
all its adjacent nodes. Let us consider again
$\mathbf{m}^{(1)}_{fv}$: upon receiving the messages
$\mathbf{m}^{(l)}_{vf}$, $l = 2,\ldots,P$, $f_k$ may evaluate the
message $\mathbf{m}^{(1)}_{fv}$ as:
\begin{equation}
{m}^{(1)}_{fv}(q) = \sum\limits_{z_2,\ldots,z_P}\left[ f_k\left(q,z_2,\ldots,z_P\right)
\prod\limits_{p=2}^{P} {m}^{(p)}_{vf}(z_p)\right] \label{eq5}
\end{equation}
for $q=0,1$.

\begin{figure}[ptb]
\begin{center}
\includegraphics[width = 8cm]{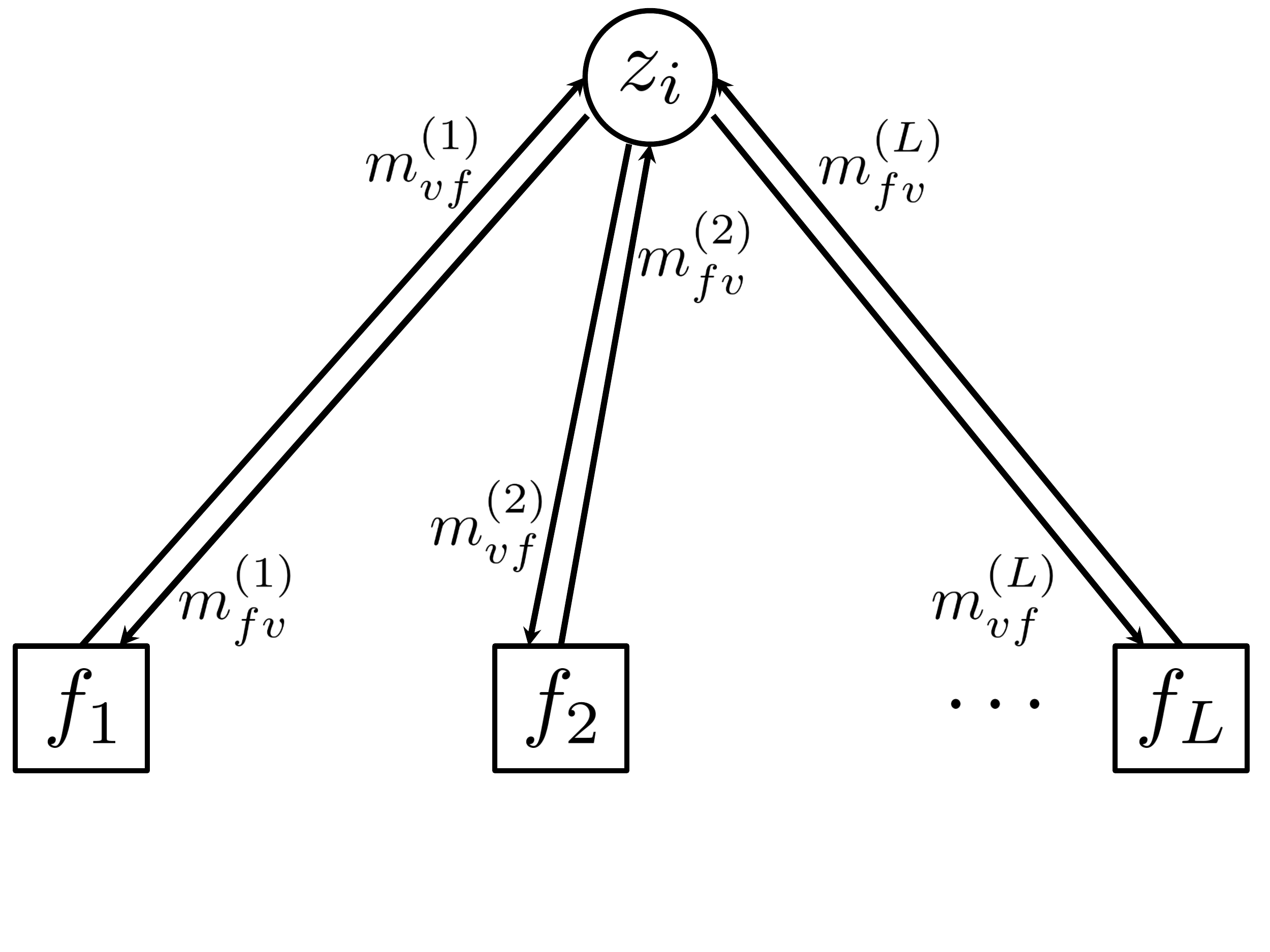}
\caption{End of message passing for node $z_i$.}%
\label{Fig3}%
\end{center}
\end{figure}

Given the message passing rules at each node, it is now possible to
derive the MP algorithm which allows to compute the marginals in (\ref{eq2}).
The process starts at
the leaf nodes, i.e., those nodes which have only one connecting
edge. In particular, each variable leaf node passes an all-ones
message to its adjacent factor node, whilst each factor leaf
node, say $f_k(z_i)$ passes the message ${m}^{(k)}_{fv}(q) =
f_k(z_i = q)$ to its adjacent node $z_i$. After initialization at
leaf nodes, for every edge we can compute the outgoing message as
soon as all incoming messages from all other edges connected to the
same node are received (according to the message passing rules
(\ref{eq4}) and (\ref{eq5})). When a message has been sent in both
directions along every edge the algorithm stops. This situation is
depicted in Fig. \ref{Fig3}: upon receiving messages from all its
adjacent factor nodes, node $z_i$ can evaluate the exact marginal as:

\begin{equation}
\mu(z_i) = \prod\limits_{k=1,\ldots,L}{m}^{(k)}_{fv}(z_i).
\label{eq6}
\end{equation}

With regard to complexity, factors to variables
message passing can be accomplished with $2^{P}$
operations, $P$ being the number of variables in $f_k$. On the other hand, variables to nodes message passing's complexity can be neglected, and, hence, the MP algorithm allows to noticeably reduce the complexity of the problem provided that the numerosity of $\mathcal{Z}_k$ is much lower than $N$. With regard to the optimization, Equation \eqref{eq6} evaluates the marginal for both $z_i = 0$ and $z_i = 1$, which represent the approximated computation of the sum-product for both hypotheses. Hence, the optimization is obtained by choosing the value of $z_i$ which maximizes it.

\subsection{Nearly-optimal data fusion by means of message passing}

\begin{figure}
\begin{center}
 \includegraphics[width=1.0\textwidth, height = 7cm]{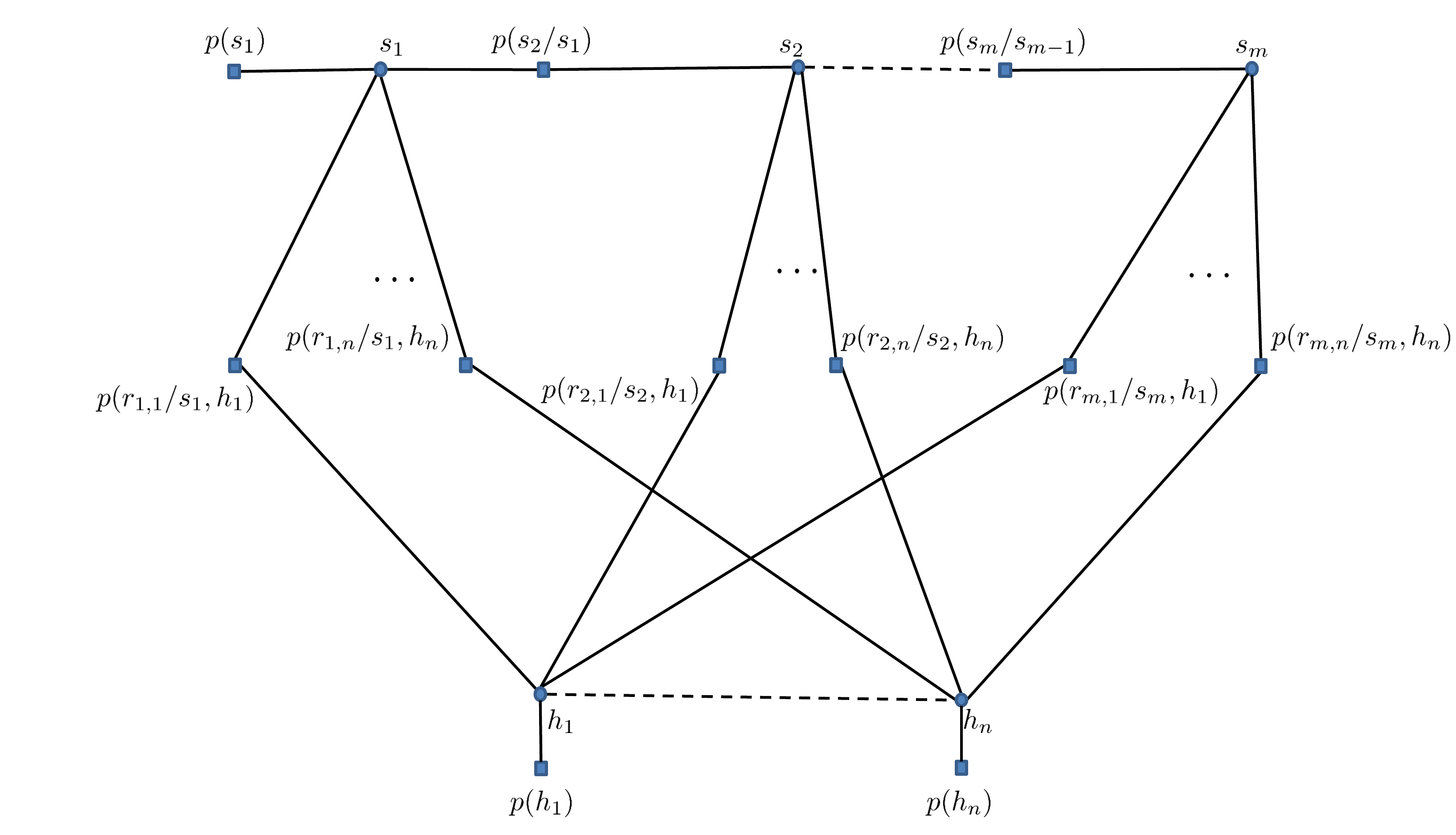} \\
\caption{Factor graph for the problem at hand.}
\label{Factor_graph0}
\end{center}
\end{figure}

The objective function of the optimal fusion rule expressed in \eqref{eqNN10} can be seen as a marginalization of a sum product of functions of binary variables, and, as such, it falls within the MP framework described in the previous Section. More specifically, in our problem, the variables are the system states $s_i$ and the status of the nodes $h_j$, while the functions are the probabilities of the reports shown in equations \eqref{eqerr1} and \eqref{eqerr2}, the conditional probabilities $p(s_i|s_{i-1})$, and the a-priori probabilities $p(h_j)$. The resulting bipartite graph is shown in Figure \ref{Factor_graph0}.

It is worth noting that the graph is a loopy graph, i.e., it contains cycles, and as such it is not a tree. However, although it was originally designed for acyclic graphical models, it was found that the MP algorithm can be used for general graphs, e.g., in channel decoding problems \cite{Rich}. In general, when the marginalization problem is associated to a loopy graph, the implementation of MP requires to establish a scheduling policy to initiate the procedure, so that variable nodes may receive messages from all the connected factors, thus evaluating the marginals. In this case, a single run of the MP algorithm may not be sufficient to achieve a good approximation of the exact marginals, and progressive refinements must be obtained through successive iterations. However, in the presence of loopy graphs, there is no guarantee of either convergence or optimality of the final solution. In many cases, the performance of the message-passing algorithms is closely related to the structure of the graph, in general, and its cycles, in particular. Many previous works in the field of channel coding, e.g., see \cite{MaoLDPC}, reached the conclusion that, for good performance, the factor graph should not contain short cycles. In our case, it is possible to see from Figure \ref{Factor_graph0} that the shortest cycles have order 6, i.e., a message before returning to the sender must cross at least six different nodes. We speculate that such a minimum cycles length is sufficient to provide good performance for the problem at hand. We will prove through simulations that such a conjecture is true.

\begin{sidewaysfigure}
\begin{center}
 \includegraphics[width=1.0\textwidth, height=13cm]{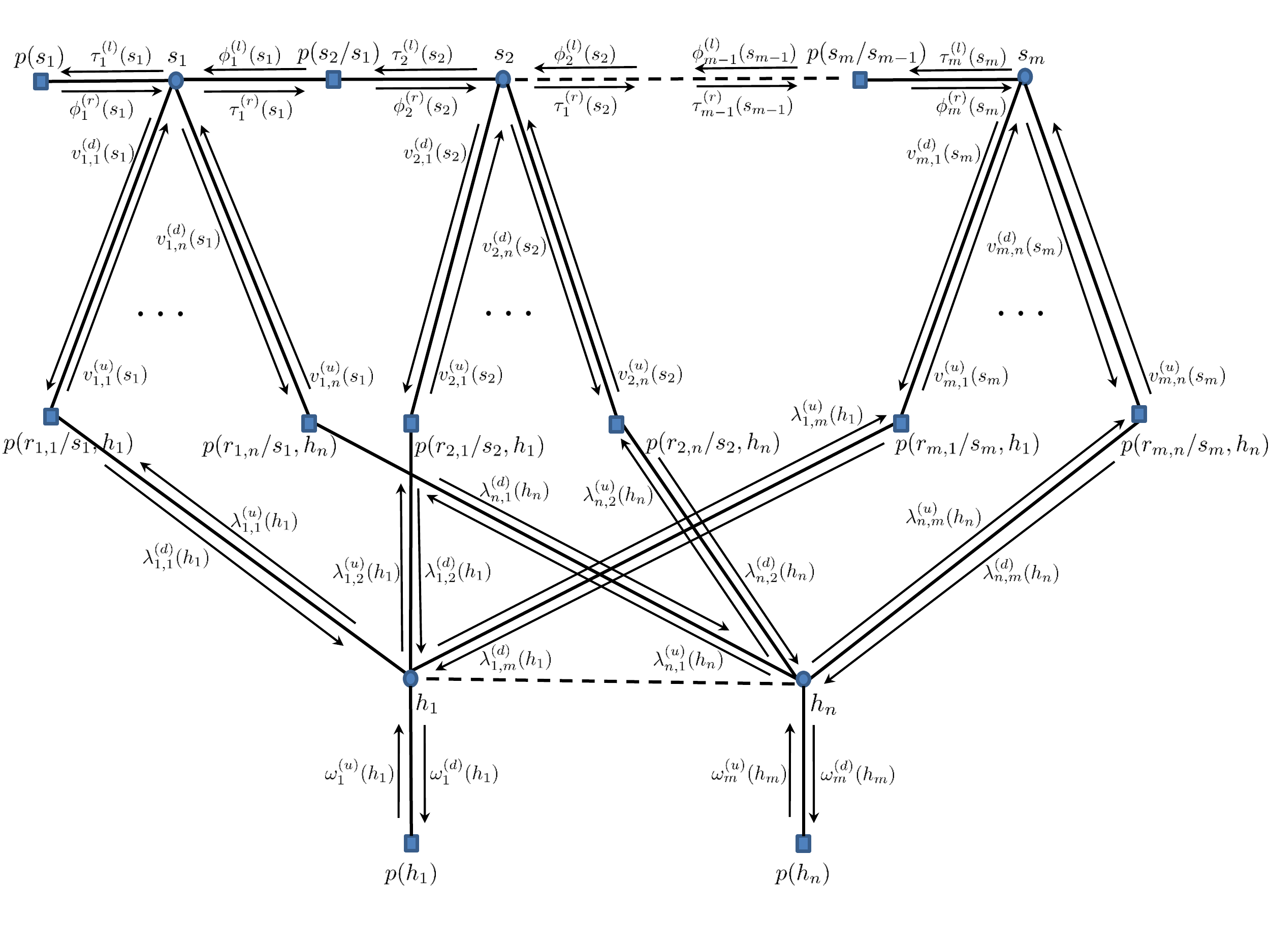} \\
\caption{Factor graph for the problem at hand with the illustration of all the exchanged messages.}
\label{Factor_graph1}
\end{center}
\end{sidewaysfigure}

To elaborate further, based on the graph of Figure \ref{Factor_graph0} and on the general MP rules reported in the previous Section, we are now capable of deriving the messages for the scenario at hand. In Figure \ref{Factor_graph1}, we display all the exchanged messages for the graph in Figure \ref{Factor_graph0} that are exchanged to estimate in parallel each of the states $s_i, i \in \{0,1\}$ in the vector $\mathbf{s} = \left\{s_1,s_2,\ldots,s_m\right\}$.
Specifically, we have:
\begin{equation}
\begin{array}{cccc}
\tau_{i}^{(l)}(s_i)  & = & \varphi_i^{(l)}(s_i) \prod\limits_{j = 1}^{n} \nu_{i,j}^{(u)}(s_i)  & i = 1,\ldots,m \\
\tau_{i}^{(r)}(s_i) & = & \varphi_i^{(r)}(s_i) \prod\limits_{j = 1}^{n} \nu_{i,j}^{(u)}(s_i)  & i = 1,\ldots,m \\
\varphi_i^{(l)}(s_i) & = &  \sum\limits_{s_{i+1}  = 0,1} p\left(s_{i+1} | s_i\right) \tau_{i+1}^{(l)}(s_{i+1}) &  i = 1,\ldots,m-1\\
\varphi_i^{(r)}(s_i) & = & \sum\limits_{s_{i-1} = 0,1} p\left(s_i | s_{i-1} \right)  \tau_{i-1}^{(r)}(s_{i-1}) &  i = 2,\ldots,m\\
\varphi_1^{(r)}(s_1) & = & p(s_1) & \\
\nu_{i,j}^{(u)}(s_i) & = &  \sum\limits_{h_j= 0,1} p\left(r_{i,j} \left| s_i,h_j\right. \right) \lambda_{j,i}^{(u)}(h_j) & i = 1,\ldots,m, ~~ j = 1,\ldots,n \\
\nu_{i,j}^{(d)}(s_i) & = & \varphi_i^{(r)}(s_i)  \varphi_i^{(l)}(s_i)  \prod \limits_{k = 1  \atop{k \ne j}}^{n}  \nu_{i,k}^{(u)}(s_i)  & i = 1,\ldots,m-1, ~~ j = 1,\ldots,n\\
\nu_{m,j}^{(d)}(s_m) & = & \varphi_i^{(r)}(s_m)  \prod \limits_{k = 1  \atop{k \ne j}}^{n}  \nu_{m,k}^{(u)}(s_m)  &  j = 1,\ldots,n\\
\lambda_{j,i}^{(d)}(h_j) & = & \sum\limits_{s_i= 0,1} p\left(r_{i,j} \left| s_i,h_j\right. \right)  \nu_{i,j}^{(d)}(s_i) & i = 1,\ldots,m, ~~ j = 1,\ldots,n\\
\lambda_{j,i}^{(u)}(h_j) & = & \omega_j^{(u)}(h_j) \prod \limits_{q = 1  \atop{q \ne i}}^{m} \lambda_{j,q}^{(d)}(h_j) & i = 1,\ldots,m, ~~ j = 1,\ldots,n\\
\omega_j^{(d)}(h_j) & = &  \prod \limits_{ i = 1 }^{m} \lambda_{j,i}^{(d)}(h_j) &  j = 1,\ldots,n\\
\omega_j^{(u)}(h_j) & = & p(h_j) &  j = 1,\ldots,n\\
\end{array}
 \label{eq_all_messages}
\end{equation}

As for the scheduling policy, we initiate the MP procedure by sending the messages $\lambda_{j,i}^{(u)}(h_j) = \omega_j^{(u)}(h_j)$ to all $p\left(r_{i,j} \left| s_i,h_j\right. \right)$ factor nodes, and by sending the message $p(s_1)$ to the variable node $s_1$. Hence, the MP proceeds according to the general message passing rules, until all variable nodes are able to compute the respective marginals. When this happens, the first iteration is concluded. Then, successive iterations are carried out by starting from leaf nodes and by taking into account the messages received at the previous iteration for the evaluation of new messages. Hence, the algorithm is stopped upon achieving convergence of messages, or after a maximum number of iterations.

The MP scheme described above can be simplified by observing that messages can be normalized without affecting the normalized marginals. Henceforward, let us consider as normalization factors the sum of the elements of the messages, i.e., if we consider for example $\tau_{i}^{(l)}(s_i)$, the normalization factor is $\tau_{i}^{(l)}(0)+\tau_{i}^{(l)}(1)$. In this case, the normalized messages, say $\bar{\tau}_{i}^{(l)}(s_i)$ can be conveniently represented as scalar terms in the interval $(0,1)$, e.g., we can consider $\bar{\tau}_{i}^{(l)}(0)$ only since $\bar{\tau}_{i}^{(l)}(1) = 1 - \bar{\tau}_{i}^{(l)}(0)$. Accordingly, the normalized messages can be evaluated as:
\begin{equation}
\begin{array}{cccc}
\bar{\tau}_{i}^{(l)}  & = & \frac{\bar{\varphi}_i^{(l)} \prod\limits_{j = 1}^{n} \bar{\nu}_{i,j}^{(u)}}{\bar{\varphi}_i^{(l)} \prod\limits_{j = 1}^{n} \bar{\nu}_{i,j}^{(u)} + (1-\bar{\varphi}_i^{(l)}) \prod\limits_{j = 1}^{n} (1-\bar{\nu}_{i,j}^{(u)})}  & i = 1,\ldots,m \\
\bar{\tau}_{i}^{(r)} & = & \frac{\bar{\varphi}_i^{(r)}  \prod\limits_{j = 1}^{n} \bar{\nu}_{i,j}^{(u)} }{\bar{\varphi}_i^{(r)} \prod\limits_{j = 1}^{n} \bar{\nu}_{i,j}^{(u)}   + (1-\bar{\varphi}_i^{(r)} ) \prod\limits_{j = 1}^{n} (1-\bar{\nu}_{i,j}^{(u)} ) } & i = 1,\ldots,m \\
\bar{\varphi}_i^{(l)} & = &  \rho \bar{\tau}_{i+1}^{(l)} + (1-\rho) (1-\bar{\tau}_{i+1}^{(l)})  &  i = 1,\ldots,m-1\\
\bar{\varphi}_i^{(r)} & = &  \rho \bar{\tau}_{i-1}^{(r)} + (1-\rho) (1-\bar{\tau}_{i-1}^{(r)}) &  i = 2,\ldots,m\\
\bar{\varphi}_1^{(r)} & = & p(s_1 = 0) & \\
\bar{\nu}_{i,j}^{(u)} & = & \frac{p\left(r_{i,j} \left| 0,0\right. \right)\bar{ \lambda}_{j,i}^{(u)} + p\left(r_{i,j} \left| 0,1\right. \right) (1-\bar{\lambda}_{j,i}^{(u)})}{p\left(r_{i,j} \left| 0,0\right. \right)\bar{ \lambda}_{j,i}^{(u)} + p\left(r_{i,j} \left| 0,1\right. \right) (1-\bar{\lambda}_{j,i}^{(u)}) + p\left(r_{i,j} \left| 1,0\right. \right)\bar{ \lambda}_{j,i}^{(u)} + p\left(r_{i,j} \left| 1,1\right. \right) (1-\bar{\lambda}_{j,i}^{(u)})} & i = 1,\ldots,m, ~~ j = 1,\ldots,n \\
\bar{\nu}_{i,j}^{(d)} & = & \frac{\bar{\varphi}_i^{(r)} \bar{ \varphi}_i^{(l)}  \prod \limits_{k = 1  \atop{k \ne j}}^{n} \bar{\nu}_{i,k}^{(u)}}{\bar{\varphi}_i^{(r)} \bar{ \varphi}_i^{(l)}  \prod \limits_{k = 1  \atop{k \ne j}}^{n} \bar{\nu}_{i,k}^{(u)} + (1-\bar{\varphi}_i^{(r)}) (1-\bar{ \varphi}_i^{(l)})  \prod \limits_{k = 1  \atop{k \ne j}}^{n} (1-\bar{\nu}_{i,k}^{(u)})}  & i = 1,\ldots,m-1, ~~ j = 1,\ldots,n\\
\bar{\nu}_{m,j}^{(d)} & = & \frac{\bar{\varphi}_m^{(r)} \prod \limits_{k = 1  \atop{k \ne j}}^{n} \bar{\nu}_{m,k}^{(u)}}{\bar{\varphi}_m^{(r)}  \prod \limits_{k = 1  \atop{k \ne j}}^{n} \bar{\nu}_{m,k}^{(u)} + (1-\bar{\varphi}_m^{(r)})   \prod \limits_{k = 1  \atop{k \ne j}}^{n} (1-\bar{\nu}_{m,k}^{(u)})}  & j = 1,\ldots,n\\
\bar{\lambda}_{j,i}^{(d)} & = & \frac{p\left(r_{i,j} \left| 0,0\right. \right)  \bar{\nu}_{i,j}^{(d)} + p\left(r_{i,j} \left| 1,0\right. \right)  (1-\bar{\nu}_{i,j}^{(d)})}{p\left(r_{i,j} \left| 0,0\right. \right)  \bar{\nu}_{i,j}^{(d)} + p\left(r_{i,j} \left| 1,0\right. \right)  (1-\bar{\nu}_{i,j}^{(d)}) + p\left(r_{i,j} \left| 0,1\right. \right)  \bar{\nu}_{i,j}^{(d)} + p\left(r_{i,j} \left| 1,1\right. \right)  (1-\bar{\nu}_{i,j}^{(d)})} & i = 1,\ldots,m, ~~ j = 1,\ldots,n\\
\bar{\lambda}_{j,i}^{(u)}  & = & \frac{\bar{\omega}_j^{(u)} \prod \limits_{q = 1  \atop{q \ne i}}^{m} \bar{\lambda}_{j,q}^{(d)} }{\bar{\omega}_j^{(u)} \prod \limits_{q = 1  \atop{q \ne i}}^{m} \bar{\lambda}_{j,q}^{(d)}  + (1-\bar{\omega}_j^{(u)}) \prod \limits_{q = 1  \atop{q \ne i}}^{m} (1-\bar{\lambda}_{j,q}^{(d)}) }& i = 1,\ldots,m, ~~ j = 1,\ldots,n\\
\bar{\omega}_j^{(d)} & = &  \frac{\prod \limits_{ i = 1 }^{m} \bar{\lambda}_{j,i}^{(d)}}{\prod \limits_{ i = 1 }^{m} \bar{\lambda}_{j,i}^{(d)}+ \prod \limits_{ i = 1 }^{m} (1-\bar{\lambda}_{j,i}^{(d)})} &  j = 1,\ldots,n\\
\bar{\omega}_j^{(u)} & = & p(h_j = 0) &  j = 1,\ldots,n\\
\end{array}
 \label{eq.eq_all_normalized_messages}
\end{equation}

\section{Simulation Results and Discussions}
\label{sec:Simulations}

In this section, we analyze the performance of the MP decision fusion algorithm. We first consider the computational complexity, then we pass to evaluate the performance in terms of error probability. In particular, we compare the performance of the MP-based scheme to those of the optimum fusion rule \cite{TIFSDFByz} (whenever possible), the soft isolation scheme presented in  \cite{CDC14}, the hard isolation scheme described in \cite{VarshneyHardIso} and the simple majority rule. In our comparison, we consider both independent and Markovian system states, for both small and large observation window $m$.

\subsection{Complexity Discussion}

In order to evaluate the complexity of the message passing algorithm and compare it to that of the optimum fusion scheme, we consider both the number of operations and the running time. By number of operations we mean the number of additions, substractions, multiplications and divisions performed by the algorithm to estimate the vector of system states $\mathbf{s}$.

\begin{figure}
\begin{center}
 \includegraphics[width=1.0\textwidth, height=6.5cm]{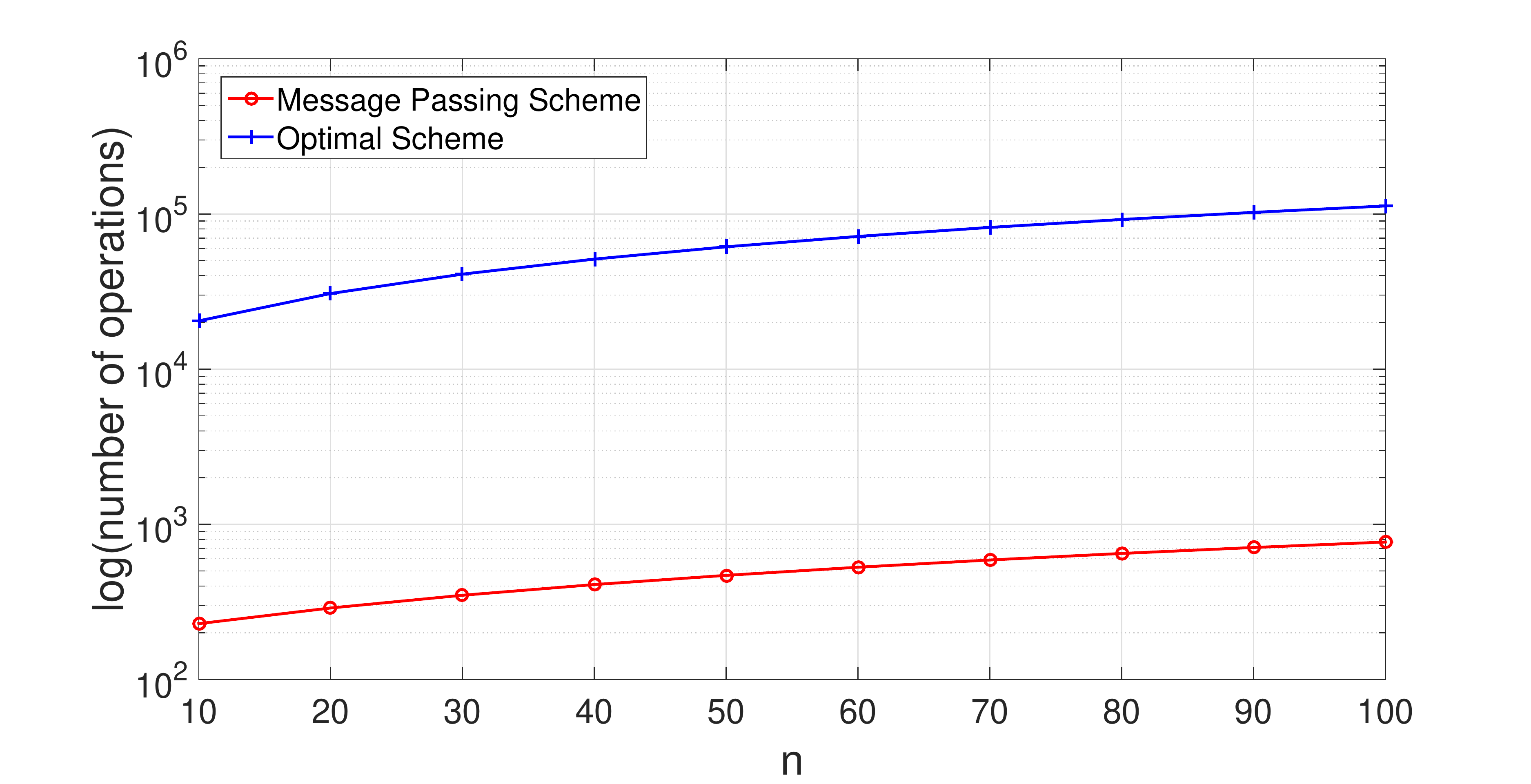} \\
\caption{Number of operations required for different $n$, $m=10$ and $5$ message passing local iterations for message passing and optimal schemes.}
\label{plot_operations_m_fixed}
\end{center}
\end{figure}

By looking at equation \eqref{eq.eq_all_normalized_messages}, we see that running the message passing algorithm requires the following number of operations:

\begin{itemize}
\item $3n+5$ operations for each of $\bar{\tau}_{i}^{(l)}$ and $\bar{\tau}_{i}^{(r)}$.

\item $3$ operations for each of $\bar{\varphi}_i^{(l)}$ and $\bar{\varphi}_i^{(r)}$.

\item $11$ operations for $\bar{\nu}_{i,j}^{(u)}$.
\item $3n+5$ operations for $\bar{\nu}_{i,j}^{(d)}$.
\item $3n+2$ operations for $\bar{\nu}_{m,j}^{(d)}$.
\item $11$ operations for $\bar{\lambda}_{j,i}^{(d)}$.
\item $3m+2$ operations for each of $\bar{\lambda}_{j,i}^{(u)}$ and $\bar{\omega}_j^{(d)}$.

\end{itemize}

\noindent summing up to $12n+6m+49$ operations for each iteration over the factor graph. On the other hand, in the case of independent node states, the optimal scheme in \cite{TIFSDFByz} requires $2^m(m+n)$ operations. Therefore, the MP algorithm is much less computationally expensive since it passes from an exponential to a linear complexity in $m$. An example of the difference in computational complexity between the optimum and the MP algorithms is depicted in Figure \ref{plot_operations_m_fixed}.


\begin{table}[h!]
\centering
\caption{Running Time (in seconds) for the Optimal and the Message Passing algorithms for: $m=10$, $\varepsilon=0.15$, $\text{Number of Trials} = 10^5$ and $\text{Message Passing Iterations} = 5$.}
\label{tab.timespent}
\renewcommand{\arraystretch}{1.1}
\begin{tabular}{c| c| c|}
\hline
\multicolumn{1}{|c|}{Setting/Scheme} & Message Passing   & Optimal      \\ \hline
\multicolumn{1}{|c|}{$n=20$,$\alpha=0.45$}  &943.807114 &1.6561e+04   \\ \hline
\multicolumn{1}{|c|}{$n=100$,$\alpha=0.49$}  &4888.821497 &2.0817e+04   \\ \hline
\end{tabular}
\end{table}

With regard to time complexity, Table \ref{tab.timespent} reports the running time of the MP and the optimal schemes. For $n=20$, the optimal scheme running time is $17.547$ times larger than that of the message passing algorithm. On the other hand, for the case of $n=100$, the optimal scheme needs around $4.258$ times more than the message passing scheme. The tests have been conducted using Matlab 2014b running on a machine with 64-bit windows 7 OS with 16,0 GB of installed RAM and Intel Core i7-2600 CPU @ 3.40GHz.

\subsection{Performance Evaluation}

In this section, we use numerical simulations to evaluate the performance of the message passing algorithm and compare them to the state of the art schemes. The results are divided into four parts. The first two parts consider, respectively, simulations performed with small and large observation windows $m$. Then, in the third part, we investigate the optimum behaviour of the Byzantines over a range of observation windows size. Finally, in the last part, we compare the case of independent and Markovian system states.

The simulations were carried out according to the following setup. We considered a network with $n = 20$ nodes, $\varepsilon = 0.15$, $\rho = \{0.95, 0.5\}$ corresponding to Markovian and independent sequence of system states, respectively. The probability $\alpha$ that a node is Byzantine is in the range $[0,0.45]$ corresponding to a number of Byzantines between 0 and 9. As to $P_{mal}$ we set it to either  0.5 or 1\footnote{It is know from \cite{TIFSDFByz} that for the Byzantines the optimum choice of $P_{mal}$ is either 0.5 or 1 depending on the considered setup.}. The number of message passing iterations is 5. For each setting, we estimated the error probability over $10^5$ trials.

\subsubsection{Small m}

\begin{figure}
\begin{center}
 \includegraphics[width=0.9\textwidth]{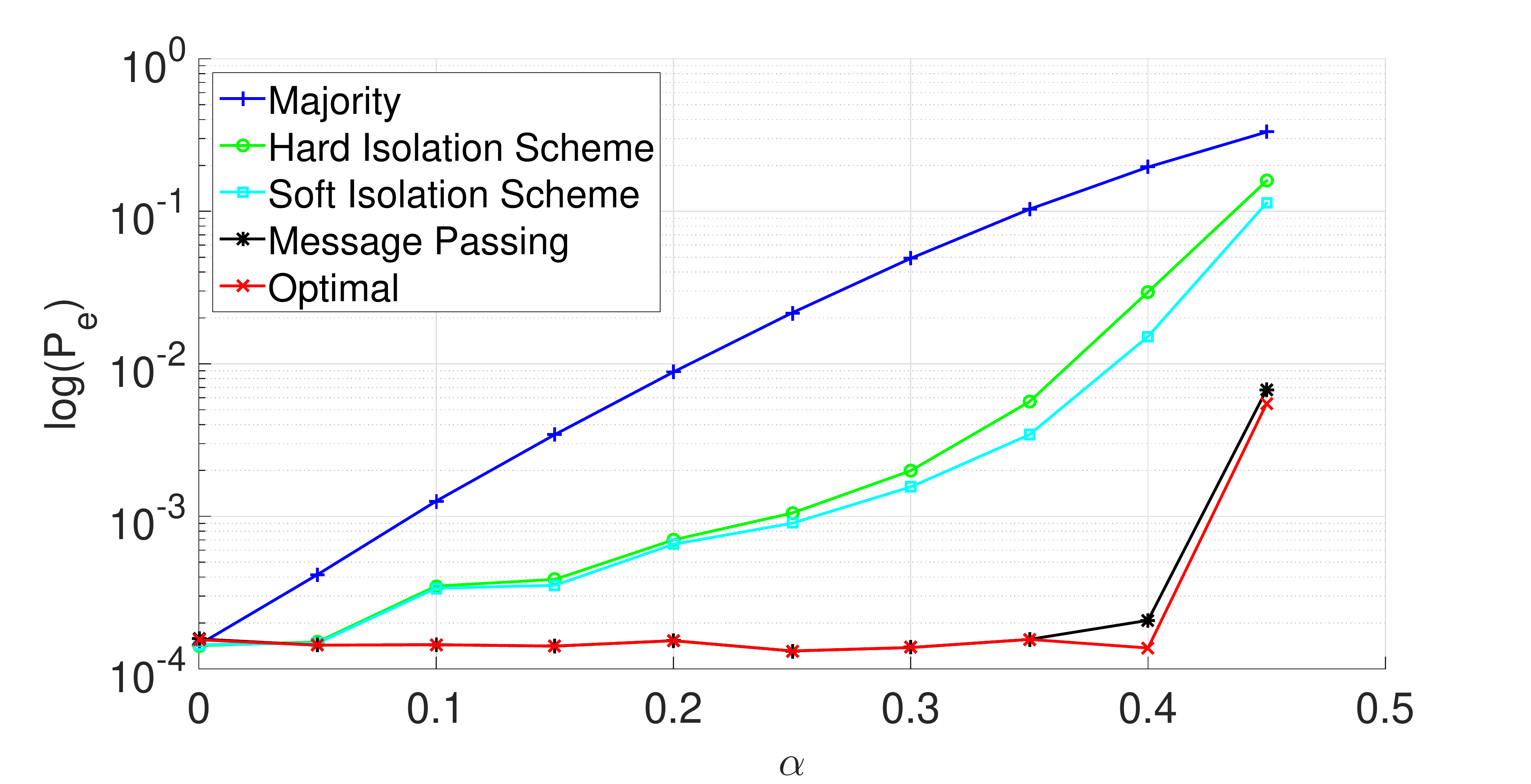} \\
\caption{Error probability as a function of $\alpha$ for the following setting: $n=20$, independent Sequence of States $\rho = 0.5$, $\varepsilon =0.15$, $m=10$ and $P_{mal}=1.0$.}
\label{Perf_one_Independent}
\end{center}
\end{figure}

\begin{figure}
\begin{center}
 \includegraphics[width=0.9\textwidth]{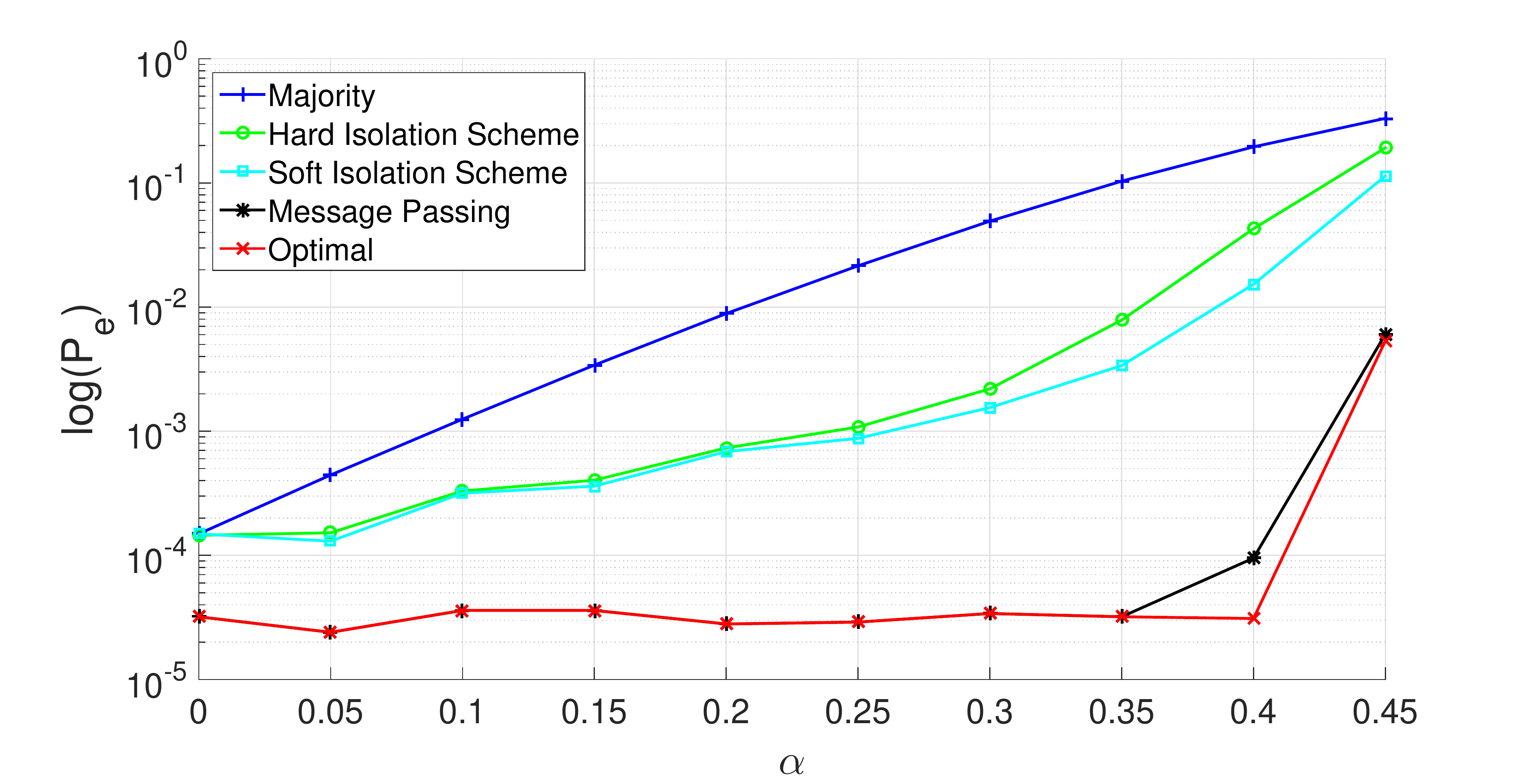} \\
\caption{Error probability as a function of $\alpha$ for the following setting: $n=20$, Markovian Sequence of States $\rho = 0.95$, $\varepsilon =0.15$, $m=10$ and $P_{mal}=1.0$}
\label{Perf_one_markov}
\end{center}
\end{figure}

To start with, we considered a small observation window, namely  $m=10$. With such a small value of $m$, in fact, it is possible to compare the performance of the message passing algorithm to that of the optimum decision fusion rule.
The results we obtained are reported in Figure \ref{Perf_one_Independent}. Upon inspection of the figure, the superior performance of the message passing algorithm over the Majority, Soft and Hard isolation schemes is confirmed. More interestingly, the message passing algorithm gives nearly optimal performance, with only a negligible performance loss with respect to the optimum scheme.

Figure \ref{Perf_one_markov} confirms the results shown in Figure \ref{Perf_one_Independent} for Markovian system states ($\rho = 0.95$).

\subsubsection{Large m}

\begin{figure}
\begin{center}
 \includegraphics[width=0.9\textwidth]{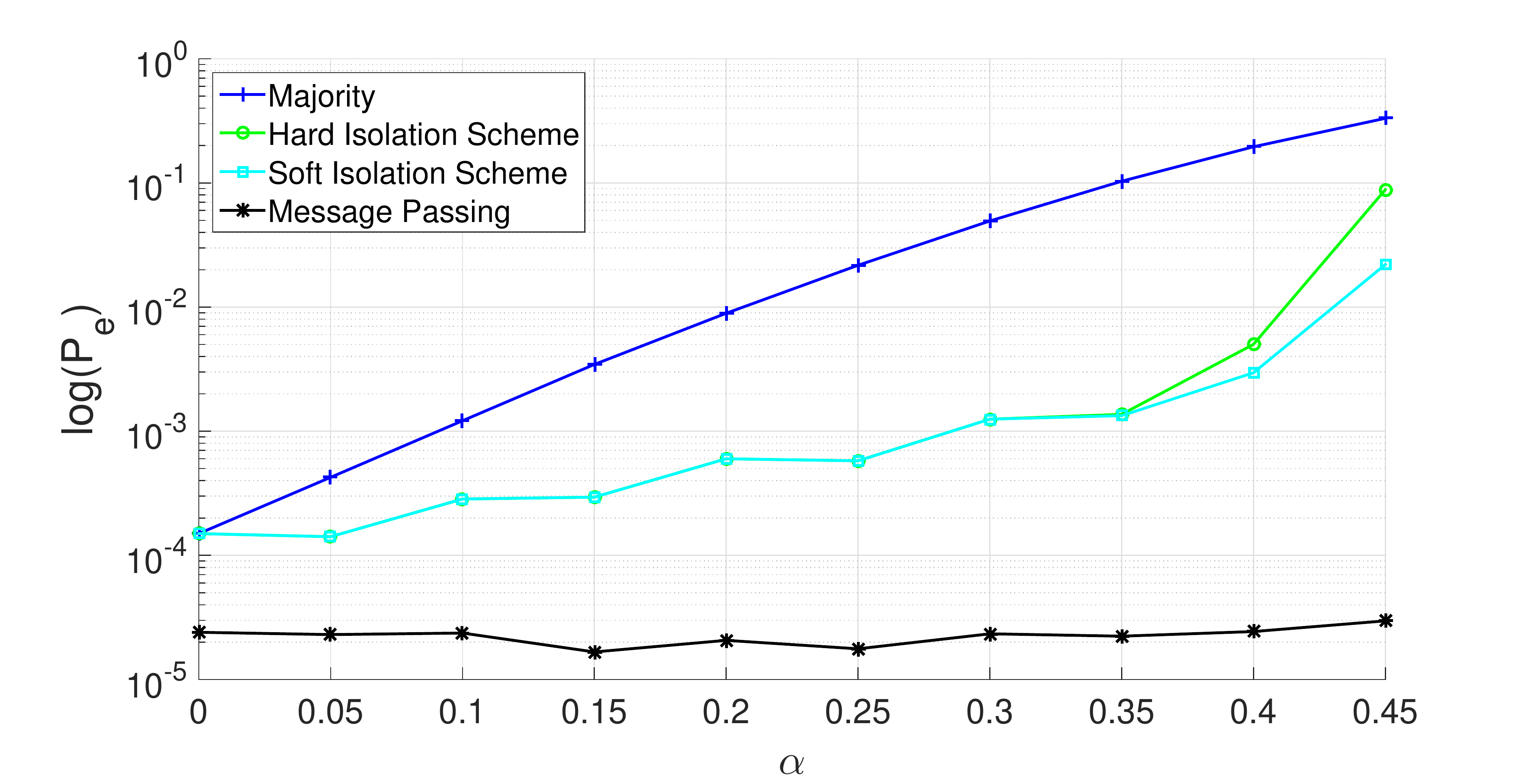} \\
\caption{Error probability as a function of $\alpha$ for the following setting: $n=20$, Markovian Sequence of States $\rho = 0.95$, $\varepsilon =0.15$, $m=30$ and $P_{mal}=1.0$.}
\label{Perf_one_markov_m_30_Pmal1}
\end{center}
\end{figure}

\begin{figure}
\begin{center}
 \includegraphics[width=0.9\textwidth]{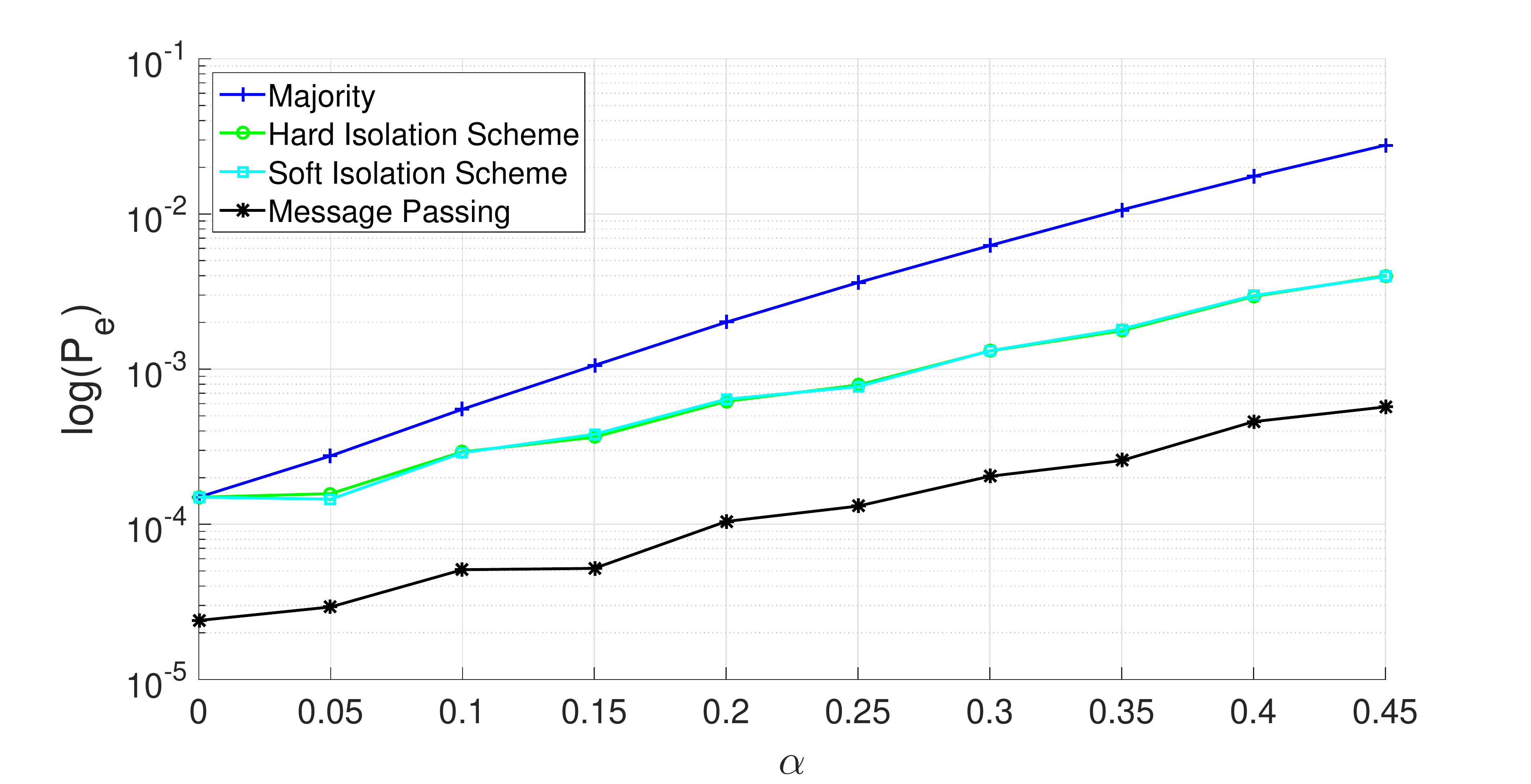} \\
\caption{Error probability as a function of $\alpha$ for the following setting: $n=20$, Markovian Sequence of States $\rho = 0.95$, $\varepsilon =0.15$, $m=30$ and $P_{mal}=0.5$.}
\label{Perf_one_markov_m_30_Pmal05}
\end{center}
\end{figure}

Having shown the near optimality of the message passing scheme for small values of $m$; we now leverage on the small computational complexity of such a scheme to evaluate its performance for large values of $m$ ($m = 30$). As shown in Figure \ref{Perf_one_markov_m_30_Pmal1}, by increasing the observation window all the schemes give better performance, with the message passing algorithm always providing the best performance. Interestingly, in this case, when the attacker uses $P_{mal}=1.0$, the message passing algorithm permits to almost nullify the attack of the Byzantines for all the values of  $\alpha$. Concerning the residual error probability, it is due to the fact that, even when there are no Byzantines in the network ($\alpha=0$), there is still an error floor caused by the local errors at the nodes $\varepsilon$. For the case of independent states, such an error floor is around $10^{-4}$. In Figure \ref{Perf_one_markov_m_30_Pmal1} and \ref{Perf_one_markov_m_30_Pmal05}, this error floor decreases to about $10^{-5}$ because of the additional a-priori information available in the Markovian case.

\subsubsection{Optimal choice of $P_{mal}$ for the Byzantines}

One of the main results proven in \cite{TIFSDFByz}, is that setting $P_{mal} = 1$ is not necessarily the optimal choice for the Byzantines. In fact, when the FC manages to identify which are the malicious nodes, it can exploit the fact the malicious nodes always flip the result of the local decision to get useful information about the system state. In such cases, it is preferable for the Byzantines to use $P_{mal} = 0.5$ since in this way the reports send to the FC does not convey any information about the status of the system. However, in \cite{TIFSDFByz}, it was not possible to derive exactly the limits determining the two different behaviours for the Byzantines due to the impossibility of applying the optimum algorithm in conjunction with large observation windows. By exploiting the low complexity of the message passing scheme, we are now able to overcome the limits of the analysis carried out in \cite{TIFSDFByz}.

\begin{figure}
\begin{center}
 \includegraphics[width=0.9\textwidth]{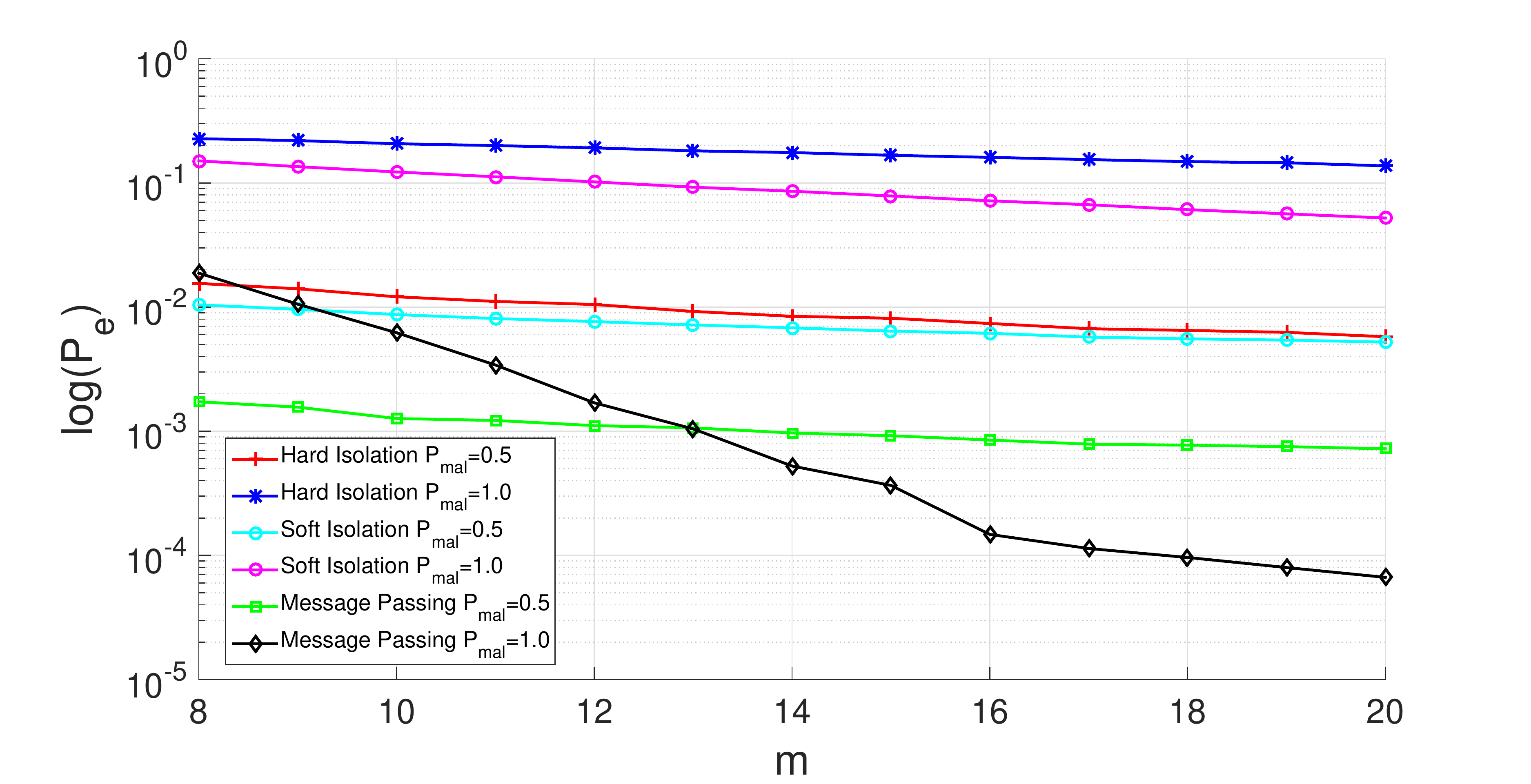} \\
\caption{Error probability as a function of $m$ for the following settings: $n=20$, Markovian Sequence of States $\rho = 0.95$, $\varepsilon =0.15$ and $\alpha=0.45$.}
\label{Perf_dual_behavior_alpha_045_markov}
\end{center}
\end{figure}

\begin{figure}
\begin{center}
 \includegraphics[width=0.9\textwidth]{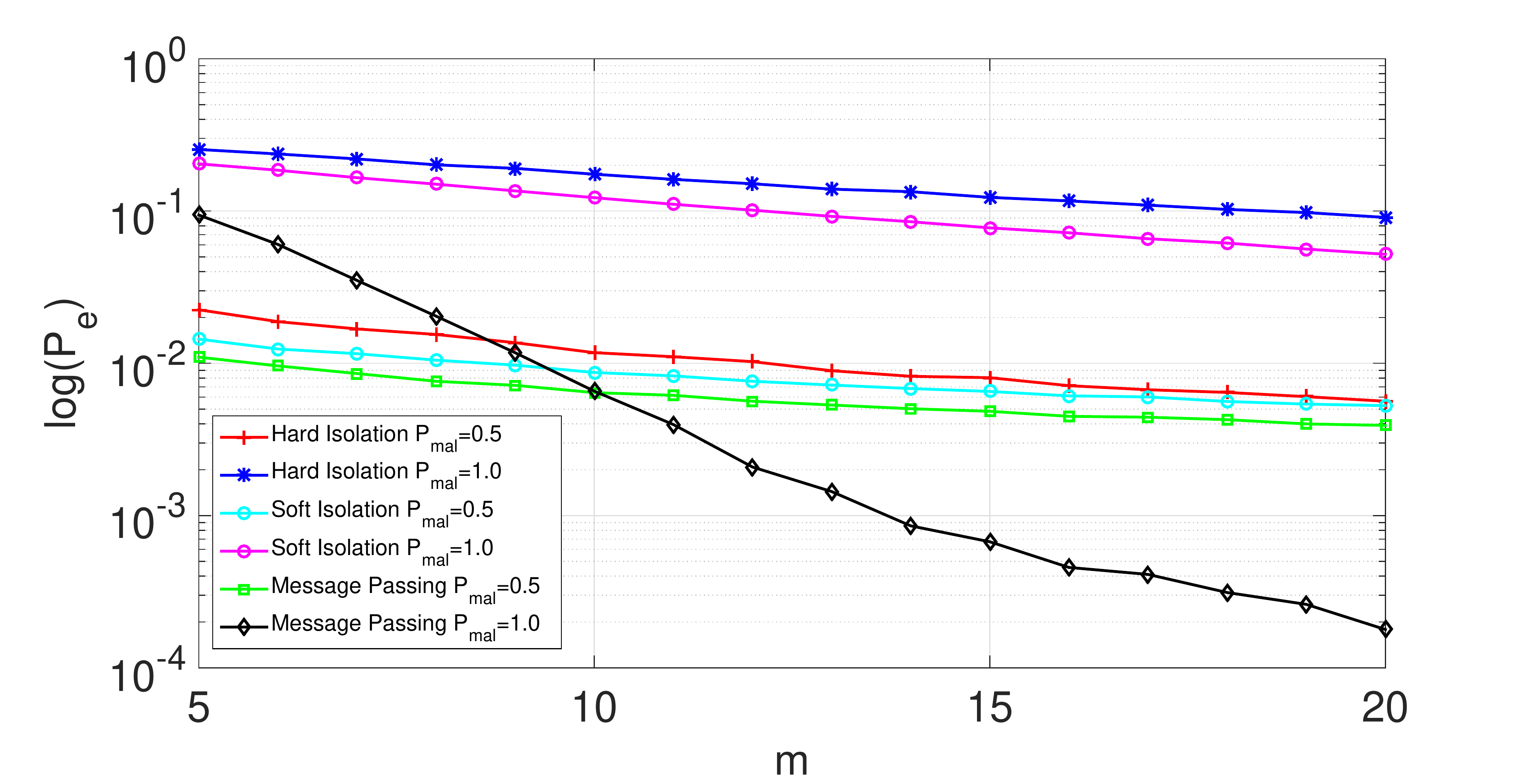} \\
\caption{Error probability as a function of $m$ for the following settings: $n=20$, independent Sequence of States $\rho = 0.5$, $\varepsilon =0.15$ and $\alpha=0.45$.}
\label{Perf_dual_behavior_alpha_045_independent}
\end{center}
\end{figure}

Specifically, we carried out an additional set of experiments by fixing $\alpha = 0.45$ and varying the observation window in the interval [5,20]. The results we obtained confirm the general behaviour observed in \cite{TIFSDFByz}. For instance, in Figure \ref{Perf_dual_behavior_alpha_045_markov}, $P_{mal} = 1.0$ remains the Byzantines' optimal choice up to $m=13$, while for $m > 13$, it is preferable for them to use $P_{mal}=0.5$. Similar results are obtained for independent system states as shown in Figure  \ref{Perf_dual_behavior_alpha_045_independent}.

\subsubsection{Comparison between independent and Markovian System States}

In this subsection, we provide a comparison between the cases of Markovian and independent system states.

By looking at Figure \ref{Perf_dual_behavior_alpha_045_markov} and \ref{Perf_dual_behavior_alpha_045_independent}, we see that the Byzantines  switch their strategy from $P_{mal}=1$ to $P_{mal} = 0.5$ for a smaller observation window ($m=10$) in the case of independent states (the switching value for the Markovian case is $m = 13$). We can explain this behaviour by observing that in the case of Markovian states, using $P_{mal} = 0.5$ results in a strong deviation from the Markovianity assumption of the reports sent to the FC thus making it easier the isolation of byzantine nodes. This is not the case with $P_{mal} = 1$, since, due to the symmetry of the adopted Markov model, such a value does not alter the expected statistics of the reports.

\begin{figure}
\begin{center}
 \includegraphics[width=0.9\textwidth]{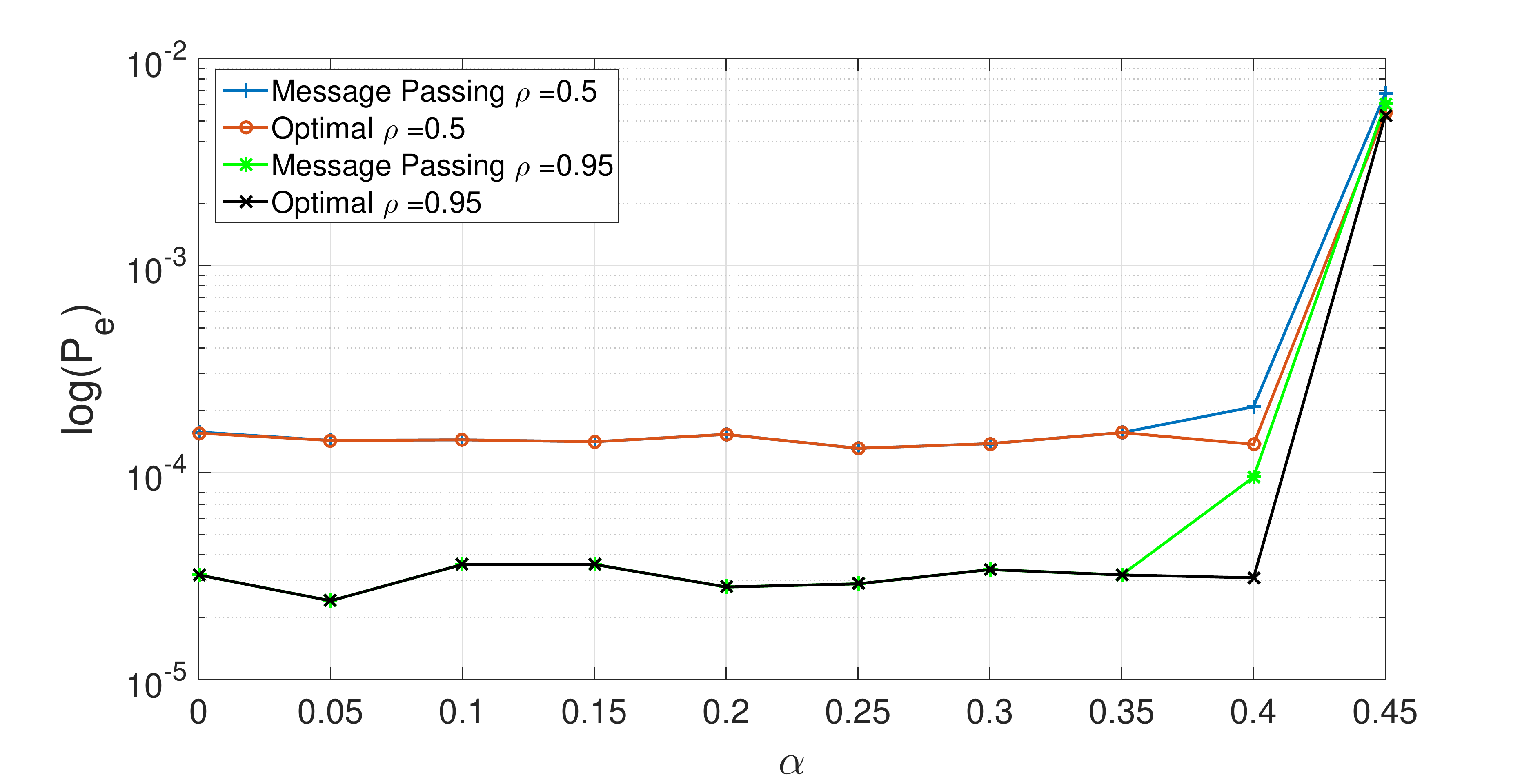} \\
\caption{Comparison between the case of independent and Markovian system states ($n=20$, $\rho = \{0.5, 0.95\}$, $\varepsilon =0.15$, $m=10$,  $P_{mal}=1.0$).}
\label{Perf_Ind_vs_Markov}
\end{center}
\end{figure}

As a last result, in Figure \ref{Perf_Ind_vs_Markov}, we compare the error probability for the case of independent and Markov sources. Since we are interested in comparing the achievable performance for the two cases, we consider only the performance obtained by the optimum and the message passing algorithms. Upon inspection of the figure, it turns out that the case of independent states is more favourable to the Byzantines than the Markov case. The reason is that the FC may exploit the additional a-priori information available in the Markov case to identify the Byzantines and hence make a better decision. Such effect disappears when $\alpha$ approaches $0.5$, since in this case the Byzantines tend to dominate the network. In that case, the Byzantines' reports prevail the pool of reports at the FC and hence, the FC becomes nearly {\em blind} so that even the additional a-priori information about the Markov model does not offer a great help.

\section{Conclusions}
\label{sec:conclusion}

In this paper, we proposed a near-optimal message passing algorithm based on factor graph for decision fusion in multi-sensor networks in the presence of Byzantines. The effectiveness of the proposed 
scheme is evaluated by means of extensive numerical simulations both for the case of independent and Markov sequence of states. 
Experiments showed that, when compared to the optimum fusion scheme, the proposed scheme permits to achieve near-optimal performance at a much lower computational cost: specifically, by adopting the new algorithm based on message passing we were able to reduce the complexity from exponential to linear. Such reduction of the complexity permits to deal with large observation windows, thus further improving the performance of the decision.
Results on large observation windows confirmed the dual behavior in the attacking strategy of the Byzantines,
looking for a trade-off between pushing the FC to make a wrong decision on one hand and reducing the mutual information between the reports and the system state on the other hand. 
In addition, the experiments showed that the case of independent states is more favorable to Byzantines than the Markovian case, due to the additional a-priori information available at the FC in the Markovian case.


As future work, we plan to focus on a scenario more favorable to the Byzantines, by giving them the possibility to access the observation vectors. In this way, they can focus their attack on the most profitable cases and avoid to flip the local decision when it is very likely that their action will have no effect on the FC decision.  Considering the case where the nodes can send to the FC more extensive reports (multi-bit case) \cite{VarshDF14} is another interesting extension.




\section*{References}

\bibliography{mybibfile}

\end{document}